\documentclass[aps,prd,preprint,tighteness,superscriptaddress,showpacs]{revtex4}
\usepackage{epsf,epsfig,graphics,graphicx,pdfpages}
\usepackage{verbatim,color,ulem}
\newcommand{\be}{\begin{equation}}
\newcommand{\ee}{\end{equation}}
\newcommand{\bea}{\begin{eqnarray}}
\newcommand{\eea}{\end{eqnarray}}
\newcommand{\ba}{\begin{array}}
\newcommand{\ea}{\end{array}}
\input epsf
\usepackage{amsmath,amssymb}
\begin{document}

\title{A Holographic Description of Negative Energy States}
\author{Da-Shin Lee}
\email{dslee@mail.ndhu.edu.tw} \affiliation{Department of Physics,
National Dong-Hwa University, Hualien, Taiwan, R.O.C.}

\author{Chen-Pin Yeh}
\email{chenpinyeh@mail.ndhu.edu.tw} \affiliation{Department of
Physics, National Dong-Hwa University, Hualien, Taiwan, R.O.C.}

\begin{abstract}
Using the AdS/CFT duality, we study the expectation value of
stress tensor in $2+1$-dimensional quantum critical theories with
a general dynamical scaling $z$, and explore various constrains on
negative energy density for strongly coupled field theories. The
holographic dual theory is the theory of gravity in
3+1-dimensional Lifshitz backgrounds. We  adopt a consistent
approach to obtain the boundary stress tensor from bulk
construction, which satisfies the trace Ward identity associated
with Lifshitz scaling symmetry. In particular, the boundary stress
tensor, constructed from the gravitational wave deformed Lifshitz
geometry, is found up to second order in gravitational wave
perturbations. {The result} is compared to its counterpart in free
{scalar} field theory at the same order in an expansion of small
squeezing parameters. This allows us to relate the boundary values
of gravitational waves to the squeezing parameters of squeezed
vacuum states. We  find that, in both cases with $z=1$, the stress
tensor satisfies the averaged null energy condition, and is
consistent with the quantum interest conjecture. Moreover, the
negative lower bound on null-contracted stress tensor, which is
averaged over time-like trajectories along nearly null directions,
is obtained. We find a weaker constraint on the magnitude and
duration of negative null energy density in strongly coupled field
theory as compared with the constraint in free relativistic field
theory. The implications are discussed.

\end{abstract}


\maketitle

\section{Introduction}
The null energy condition states that the energy-momentum  tensor
$T_{a b}$, contracted by all null vector $K_{a}$,  cannot be
negative, namely $T_{a b} K^{a} K^{b} \ge 0$. This pointwise
energy condition bounds  stress tensor in each spacetime
points, and is satisfied by most of classical matter fields.
The null energy condition ensures that light rays are focused,
and plays an essential role in the singularity theorem and some
other theorems in general relativity~\cite{PENROSE_65}. However,
it has been realized that quantum field theory allows to violate
the pointwise energy condition~\cite{EP}.   The existence of  negative
energy density might result in violation of the second law of
thermodynamics~\cite{FO,DA} and cosmic censorship~\cite{FOR}, and
can also lead to a spacetime with exotic features such as
wormholes, superluminal travel, or construction of time
machines~\cite{MO,AL,FOR_96,VISSER_98,ORI,Myrzakulov}.
Although  pointwise energy
can be negative, it is also found that any negative energy density must be
accompanied by positive energy density to place limits on the extent of the energy condition
breakdown~\cite{For_95,FO1,PF,FE}. This can be seen by considering
the average of the expectation value of the stress tensor
 on time-like or null-like geodesics. One example is the
averaged null energy condition
  \be
  \label{ANEC2}
  \int \langle T_{a b}(\lambda)\rangle K^{a} K^{b} d\lambda \ge 0
  \ee
where the average is over a complete null geodesic and $K^{a}$ is
a tangent vector to the path. {{While the averaged energy condition
is verified in wide varieties of theories and spacetime
backgrounds, some negative lower bounds are also obtained in many
cases with appropriate sampling functions.
These bounds usually
imply the existence of} quantum inequalities that limit the
magnitude and duration of negative energy density \cite{For_95}.
However for the averaged null energy condition,  the corresponding
quantum inequality, which is invariant by rescaling of an affine
parameter, is only found  in 1+1-dimensional quantum field
theory~\cite{For_95,FE_03}. Even so, it does not mean that in
higher dimensions the magnitude of negative null-contracted stress
tensor is completely unconstrained. One can consider an
alternative way to do the average, which is along a time-like
trajectory
   \be  \label{anec} \int \, g_{\tau_0} (\tau) \,  \langle T_{a
   b}(\tau)
\rangle K^{a} K^{b}  \, d \tau \, ,
   \ee
where the sampling function $g_{\tau_0} (\tau)$ is introduced, and
is a function of the proper time $\tau$ that parameterizes the
time-like trajectory. This sampling function introduces a peaked
value and a characteristic sampling scale $\tau_0$ over the proper
time $\tau$. In this case~(\ref{anec}), the quantum inequality for
null-contracted stress tensor can be constructed~\cite{FE_03}.
{According to~\cite{Ford_99}, the quantum interest conjecture
states that an energy loan, namely the negative energy, must
always be repaid by the positive energy, with an interest which is
determined by the magnitude and duration of the loan. This results
in the uncertainty-principle-type inequalities to constrain the
extent  of negative energy density. In particular, the proof of
the conjecture for free massless scalar fields in an arbitrary
quantum state in two and four-dimensional flat spacetime is
discussed in~\cite{Ford_99}. Another way to see the negative
energy density for quantum fields is to consider their subvacuum
fluctuation effects on the dynamics of a particle, with which
quantum fields are coupled~\cite{HWL,Lee_12}.  {An example of}
subvacuum phenomenon is the suppression of quantum decoherence of
a particle state due to an interaction with environmental quantum
fields with negative energy density~\cite{HS,Lee_06}. One of the
quantum states in quantum field theory that can have negative
energy density in some spacetime region is the squeezed vacuum
state~\cite{Lee_12}. The squeezed vacuum states presumably can be
generated in laboratory experiments via the nonlinear-optics
technique by squeezing the normal vacuum state, in which the
time-translational invariance is broken so as to produce
nonstationary quantum correlations. Some sort of quantum
inequality on the dynamics of the particle analogous to that for
the energy density is considered~\cite{Lee_12}. } Nevertheless
quantum inequalities are mostly studied in free field theories.
The idea of this paper is to pursue above mentioned bounds and
subsequently derive quantum inequalities in strongly coupled
fields, obtained from the holographic approach.

The holographic duality, in its original formulation, is to relate
a 4-dimensional Conformal Field Theory (CFT) to the string theory
in 5-dimensional anti-de Sitter (AdS) space~\cite{AdSCFT}. Later,
the idea of this holographic duality has been extended to other
systems such as strong coupling problems in condensed matter
physics and the hydrodynamics of the quark-gluon plasma. Moreover,
considerable efforts have been focused on using the holography
idea to explore  Brownian motion of a particle moving in a
strongly coupled
environment~\cite{Herzog:2006gh,Gubser_06,Teaney_06,Son:2009vu,Giecold:2009cg,CasalderreySolana:2009rm,Huot_2011,Holographic
QBM,Tong_12,yeh_14,yeh_15,yeh_16}. In addition, there have been
extensive studies of above mentioned energy conditions with the
holographic approach. The conjecture of the quantum null energy
condition with a lower bound related to the von Neumann entropy is
proposed to generalize the local null energy
condition~\cite{bousso_16}. This conjecture can be proved within
the holographic framework  by finding the connection between the
von Neumann entropy and the minimal surface in a gravity
background~\cite{koeller_15}. The proof of averaged null energy
conditions for a class of strongly coupled conformal field
theories is also studied within the context of AdS/CFT
correspondence \cite{kelly_14}, where the bulk causality can give
constraints on the extent of negative energy in boundary field
theories. It will be thus of  great interest  to explore these
energy bounds in the specific holographic setup. {In
\cite{Lenny_99}, the holographic stress tensor, obtained from
gravitational wave deformed $AdS_5$ spacetime, is studied and
found to behave like the one for squeezed vacuum states in free
scalar field theory up to  second order in an expansion of small
squeezing parameters. In this case the stress tensor satisfies the
averaged null energy condition, and is consistent with the quantum
interest conjecture proposed in \cite{Ford_99}.  In \cite{yeh_16},
we explore subvacuum phenomena of a probed particle coupled to the
squeezed vacuum of  strongly coupled quantum critical fields with
a dynamical scaling $z$. The holographic description corresponds
to a string moving in $4+1$-dimensional Lifshitz geometry with
gravitational wave perturbations. Additionally, the dynamics of a
probed particle is realized by the motion of the endpoint of a
string at the boundary. In this paper, we extend the study
of~\cite{Lenny_99} to quantum critical theories with their dual
gravity theory in the Lifshitz backgrounds. In particular, the
holographic stress tensor, constructed from gravitational wave
deformed Lifshitz spacetime, is studied up to second order in
gravitational wave perturbations. We find that the leading term
can have negative energy density, and in general shows oscillatory
behavior in time, while the subleading term is positive and
constant in time. It is of interest to compare with the
expectation value of stress tensor of squeezed vacuum states for
free scalar fields in a small squeezing parameter expansion. The
squeezed states are constructed by squeezing the normal vacuum
states so that the time-translational invariance is broken. Thus,
the corresponding energy-momentum tensor may become
time-dependent, and, as will be seen below, can have the same
behavior as the holographic stress tensor order by order in the
small perturbations of gravitational waves. We then study the
averaged null energy conditions, and derive the associated quantum
inequalities, if exist, for strongly coupled quantum critical
theories in the case of $z=1$ with Lorentz symmetry.

In next section, we introduce the Lifshitz geometry in $d+1$
dimensions, which is the gravity background dual to quantum
critical theories with a general dynamical scaling $z$ in
spacetime dimension $d$. The AdS/CFT prescription is adopted to
obtain the boundary stress tensor from both gravitational wave
deformed Lifshitz geometry and Lifshitz black hole in 3+1
dimensions. In Sec. 3, by comparing the stress tensor of boundary
fields to that of free relativistic fields in squeezed vacuum
states in 2+1 dimensions, the boundary values of gravitational
waves can be identified with the squeezing parameters in quantum
field theory. In Sec. 4, we show that, for small squeezing, in both
strongly coupled quantum critical field theories with $z=1$ and
free relativistic field theories the obtained stress tensor
satisfies the averaged null energy condition, and is consistent
with the quantum interest conjecture. A quantum inequality, which
constrains the extent of negative null energy density, can be
found in strongly coupled field theories. This is then to compare
with similar constraint derived from free relativistic fields. We conclude in Sec. 5. The sign convention $(-,+,+,...)$
is adopted in the $d+1$-dimension metric in dual gravity theory
with indices $\mu, \nu,..$. Indices $a,b,c...$ denote all
spacetime coordinates in boundary field theory while $i,j,k...$
denote only spatial dimensions.

\section{Lifshitz Geometry and Energy-Momentum Tensor}

The theory of  quantum critical points is  a fixed point theory with
the scaling symmetry:
  \be \label{lif_scaling}
  t\rightarrow\mu^zt \, ,\qquad\qquad x\rightarrow\mu x \, .
  \ee
The holographic dual for such quantum critical theories in
2+1-dimension has been proposed in~\cite{Kachru_08}, where the
gravity theory is in the 3+1-dimensional Lifshitz background. We start from the
general d+1-dimensional Lifshitz background with the metric,
   \be
   \label{bmetric}
   ds^2=g^{(0)}_{\mu\nu}dx^{\mu}dx^{\nu}=-\frac{r^{2z}}{L^{2z}}dt^2+\frac{L^2}{r^2}dr^2+\frac{r^2}{L^2}dx_idx_i
   \, ,
   \ee
where the above scaling symmetry~(\ref{lif_scaling}) is realized as an isometry of the
metric. This gravity background~(\ref{bmetric}) can be engineered by coupling
gravitation fields with negative cosmological constant to
massive Abelian vector fields~\cite{Marika_08}. The corresponding
action is given by:
  \be
  \label{action}
  S=\frac1{16\pi G_{d+1}}\int
  d^{d+1}x \,\sqrt{-g}\, ( R+ 2\Lambda-\frac14 {\cal F}^{\mu\nu}{\cal F}_{\mu\nu}-\frac1 2m^2 {\cal A}^{\mu} {\cal
  A}_{\mu})\,.
  \ee
In addition to the Einstein-Hilbert action and the cosmological
constant $\Lambda$ term, the action for a vector field ${\cal
A}_{\mu}$ with mass $m$ is introduced. Moreover ${\cal
F}_{\mu\nu}$ is the field strength of ${\cal A}_{\mu}$. The action
yields the equations of motion for the metric and vector fields,
   \bea \label{EoM}
   &&R_{\mu\nu}=-\frac{2 \Lambda}{d-1} g_{\mu\nu}+\frac12 g^{\alpha\beta}{\cal F}_{\mu\alpha} {\cal F}_{\nu\beta}+\frac12 m^2 {\cal A}_{\mu} {\cal A}_{\nu}-\frac1{4(d-1)} {\cal F}_{\alpha\beta} {\cal F}^{\beta\alpha}g_{\mu\nu}\, ,\\
   &&D_{\mu}{\cal F}^{\mu\nu}=m^2 {\cal A}^{\nu} \, ,
   \eea
where $D_{\mu}$ is a covariant derivative with respect to the
background metric $g_{\mu \nu}$. The vector field is assumed to be
  \be \label{A_field}
  {\cal A}_{\mu}={\cal A} \frac{r^z}{L^z}\delta^{0}_{\mu} \, .
  \ee
Then the Lifshitz background in (\ref{bmetric}) can be achieved by
setting
 \be
 \label{para}
 {\cal A}=\sqrt{\frac{2(z-1)}{z}},~~m^2=\frac{(d-1)z}{L^2},~~\Lambda=\frac{(d-1)^2+(d-2)z-z^2}{2L^2}
 \, .
 \ee
Later we will construct the boundary stress tensor from the
perturbations of gravitational waves in the Lifshitz background.

In AdS space, the boundary stress tensor can be defined by varying
the action with respect to the induced boundary metric
$\gamma_{ab}$, constructed from \be \label{metric_induced}
\gamma_{\mu\nu}=g_{\mu\nu}-n_{\mu}n_{\nu}\, , \ee
 where the unit vector $n_{\nu}$ is orthogonal to the boundary
and outward-directed~\cite{Kraus_99,Myers_99}. However, for the
Lifshitz background, which involves massive vector fields, it is
found that the boundary stress tensor is more appropriately
constructed by introducing a set of frame fields, by which the
induced metric $\gamma_{ab}$ on the boundary can be expressed in
terms of the flat metric $\eta_{AB}$ to be~\cite{Marolf_99} \be
   \gamma_{ab}=\eta_{AB} \, e_a^A  e_b^B \, .
   \ee
The introduction of the above frame fields also allows us to write
the vector fields as
\be {\cal A}_a= {\cal A}_A \, e_a^A \, . \ee
Then the conserved boundary stress tensor can be derived by
\be
  \tau^{ab}=-\frac1{\sqrt{-\gamma}}\frac{\delta S_{graity}}{\delta
  e_{a}^{A}}e^{b A} \, , \label{boundary_stress_tensor}
  \ee
where the functional derivative is taken with respect to the
on-shell gravity action $S_{gravity}$ by holding $\eta_{AB}$ and
${\cal A}_A$ fixed~\cite{Marolf_99}. The stress tensor
above~(\ref{boundary_stress_tensor}) will be evaluated at the
boundary, $r=r_b$. The stress tensor defined in this way normally
diverges when taking the boundary to infinity, $r_b \rightarrow
\infty$. To render it finite, the appropriate counterterms need to
be introduced for cancelling these divergent pieces. Although
there does not exist relativistic covariant stress tensor for a
general dynamical scaling $z$, the resulting stress tensor obeys
the trace Ward identity, which is required from the symmetry of
Lifshitz scaling~\cite{Marika_08}. In below, we will mainly
consider the 3+1-dimensional Lifshitz background, and the full
action (\ref{action}) in 3+1 dimensions becomes
  \be
  \label{4d action}
  S_{3+1}=\frac1{16\pi G_4}\int
  d^{4}x  \sqrt{-g} \, ( R + 2\Lambda-\frac14 {\cal F}^{\mu\nu}{\cal F}_{\mu\nu}-\frac12 m^2 {\cal A}^{\mu} {\cal
  A}_{\mu}) \, ,
  \ee
 and the counterterms are found uniquely to be \cite{Ross_09}
  \be
  S_c=\frac1{8\pi G_4}\int
  d^{3}\, \zeta\sqrt{-\gamma} \big( \Theta-2-\frac{z}2 {\cal A} \sqrt{-{\cal A}^{a} {\cal A}_{a}}
  \big) \, ,
  \ee
where $\zeta^a$ are the coordinates on the boundary, and ${\cal A}$
is defined in (\ref{para}) for $d=3$. In addition, $\Theta$ is the
trace of the extrinsic curvature on the boundary,
  \be \label{ex_curvature}
\Theta_{\mu\nu}=-\gamma_{\mu}^{\alpha}D_{\alpha}n_{\nu} \, ,
  \ee
in which the unit vector $n_{\nu}$ again is orthogonal to the boundary
and outward-directed.
We thus consider the gravity action as $S_{gravity}=S_{3+1}+S_C$
and choose the coordinates so that the background is
asymptotically with the metric $g^{(0)}_{\mu\nu}$ in
(\ref{bmetric}). {With the appropriately chosen
frame fields $e^0_a=(\frac{r}{L})^z\delta^0_a$ and
$e^i_a=\frac{r}{L}\delta^i_a$, the definition of the energy-momentum tensor in~(\ref{boundary_stress_tensor}) gives
\cite{Ross_09}}
  \be
  \label{stress}
  16\pi G_4 \tau_{00}=2S_{00}+ S_{0} {\cal A}_{0},~~~16\pi G_4 \tau_{ij}=2 S_{ij}+ S_{i} {\cal
  A}_{j} \, ,
  \ee
  \be
  \label{stress2}
  16\pi G_4 \tau_{0i}=-2S_{0i}- S_{0} {\cal A}_{i},~~~16\pi G_4 \tau_{i0}=-2 S_{i0}- S_{i} {\cal
  A}_{0} \, ,
  \ee
 where
   \be
   S_{ab}=\Theta_{ab}-\gamma_{ab}\Theta-\frac2{L}\gamma_{ab}-
   \frac{z}{2L}\frac{{\cal A}}{\sqrt{-{\cal A}_c {\cal A}^c}}( {\cal A}_a {\cal A}_b-{\cal A}_c {\cal A}^c\gamma_{ab})
   \ee
and
   \be
   S_{a}=-{\cal F}_{a\nu}n^{\nu}+\frac{z}L \frac{{\cal A}}{\sqrt{-{\cal A}_c {\cal A}^c}}{\cal A}_{a}
   \, .
   \ee
Because $S_a$ and ${\cal A}_a$ have the time component only and
$S_{ab}$ is diagonal, the components of $\tau_{0i}$ and
$\tau_{i0}$ are zero. Therefore the nonzero components $\tau_{00}$
and $\tau_{ij}$ will be computed later. The boundary fields live
on the flat metric $\eta_{ab}$, which is related to the induced
metric $\gamma_{ab}$ by the conformal transformation. Thus the
expectation values of the stress tensor operators, $\langle
T_{ab}\rangle$ can be derived from $\tau_{ab}$ by
   \be
     \label{cfactor}
   \sqrt{-\eta}\eta^{ab}\langle
   T_{bc}\rangle=\sqrt{-\gamma}\gamma^{ab}\tau_{bc} \, .
   \ee
In what follows, we will consider the gravitation perturbations in
dual gravity theory, from which the stress tensor of boundary
fields is obtained.

\subsection{Gravitational Waves}
The above-mentioned method will be adopted to calculate the
boundary stress tenor dual to the Lifshitz background with the
perturbations from gravitational waves. We consider the background
with the metric $g^{(0)}_{\mu\nu}$ in (\ref{bmetric}) plus small
perturbations due to gravitational waves,
   \be
   \label{gravitonp}
   g_{\mu\nu}=g^{(0)}_{\mu\nu}+\delta g^{(1)}_{\mu\nu} \, .
   \ee
These gravitational waves can be parameterized as
  \be
  \label{gravitonp2}
  \delta g^{(1)}_{\mu\nu}=\xi_{\mu\nu}\frac{r^2}{L^2}\, \phi(t,r) \, ,
  \ee
where $\xi_{\mu\nu}$ is a traceless, symmetric constant tensor
with nonzero components $\xi_{i\,j}$ along spatial directions
$i,j$ only. The equation of motion for gravitational waves can be
found by linearizing~(\ref{EoM}) around the background solutions
in~(\ref{bmetric}) and~({\ref{A_field}) given by
   \be
   \label{eom}
   -\frac{r^{-2z}}{L^{-2z-2}}\, \partial_t^2\phi(t,r)+(3+z)\, r\partial_r\phi(t,r)+r^2\, \partial_r^2\phi(t,r)=0\, . \ee
The normalizable general solution can be written as
  \be
  \label{graviton}
  \phi(t,r)=r^{-\frac{2+z}{2}}\int_0^{\infty}
  d\omega\, \varphi(\omega) \, J_{\frac{2+z}{2z}}\, \bigg(\frac{L^{z+1}\omega}{zr^z}\bigg)\, e^{-i\omega
  t}+{\rm h. c.} \, ,
  \ee
where $J_{\nu}(x)$ is the Bessel function of the first kind. The
boundary is the time-like surface located at $r=r_b$ with normal
unit vector $n^{\mu}=\frac{r_b}{L}\delta^{\mu}_r$. The induced
boundary metric $\gamma_{ab}$ defined in~(\ref{metric_induced})
then becomes
   \be
 \gamma_{00}=-\frac{r_b^{2z}}{L^{2z}},~~~~\gamma_{ij}=\frac{r_b^2}{L^2}\delta_{ij}+\xi_{ij}\frac{r_b^2}{L^2}\phi(t,r_b)
  \, . \ee
To linear order in gravitational perturbations, the extrinsic
curvature  (\ref{ex_curvature}) is obtained as
  \be
\Theta_{00}=\frac{r_b^{2z}z}{L^{2z+1}},~~~\Theta_{ij}=-\frac{r_b^2}{L^3}\delta_{ij}-\xi_{ij}\frac{r_b^2}{L^3}\left(
\phi(t,r_b)+\frac{r_b}{2}\partial_r\phi(t,r_b)\right) \, ,
  \ee
and from (\ref{stress}) the boundary stress tensor can be read off
to be
  \be
  16\pi G_4\tau_{00}=0,~~~16 \pi
  G_4\tau_{ij}=-\xi_{ij}\frac{r_b^3}{L^3}\, \partial_r\phi(t,r_b) \, .
  \ee
Thus, using the conformal factor (\ref{cfactor}), the expectation
value of the stress tensor which is $r_b$-independent in the
$r_b\rightarrow\infty$ limit, is found, to leading order in
gravitational wave perturbations, to be
  \be
  \label{psi}
  16\pi G_4\langle
  T^{(1)}_{00}\rangle =\frac{r_b^{2-z}}{L^{2-z}}\tau_{00}=0,~~~16\pi G_4\langle
  T^{(1)}_{ij}\rangle=\frac{r_b^{z}}{L^{z}}\tau_{ij}=-\frac{r_b^{3+z}}{L^{3+z}}\, \xi_{ij}\, \partial_r\phi(t,r_b)\,
  \ee
To see this, we use the asymptotic property of the Bessel function
$J_{\nu}(x\rightarrow0)\approx x^{\nu}$ and the behavior of $\phi$
in (\ref{graviton}) is found to be
 \be
 \label{bc}
 \phi(t,r\rightarrow\infty)= r^{-2-z} \, \int d\omega \, \varphi(\omega)\frac1{\Gamma(\frac1z+\frac32)}\bigg(\frac{L^{z+1}\omega}{2z}\bigg)^{\frac{2+z}{2z}}\, e^{-i\omega t}\, .
 \ee
Substituting the above form into (\ref{psi}),  $\langle
\hat{T}_{ij}(t,r_b\rightarrow\infty)\rangle$ is independent of
$r_b$, namely
  \be
  \label{st1}
  \langle T^{(1)}_{ij}(t,r_b\rightarrow\infty)\rangle \equiv \langle T^{(P)}_{ij} (t) \rangle =\frac{L^{\frac1{z}-\frac32-\frac{z}2}}{16\pi G_4}\, \,
  \frac{(2+z)(2z)^{-\frac{2+z}{2z}}}{\Gamma(\frac1z+\frac32)}\, \xi_{ij}\int
  d\omega \, \varphi(\omega) \, \omega^{\frac{2+z}{2z}}\, e^{-i\omega t}
  \, ,
  \ee
where $\varphi(\omega)$ will be fixed later by the boundary
condition of gravitational waves in (\ref{bc}). As long as
$\varphi(\omega \rightarrow 0) $ is regular, $ \langle
T^{(P)}_{ij}(t)\rangle $ vanishes as $t\rightarrow \infty$. As
in~\cite{Lenny_99}, we assume that the gravitational waves are
generated at time $t=0$. They then propagate toward large $r$, and
reach the boundary. As a result, the boundary stress tensor
reveals an oscillatory behavior in time and can be negative,
leading to the so-called energy loan.  In the terminology
introduced by \cite{Ford_99}, this contribution to the expectation
value of the stress tensor  is called the {\it principal}, denoted
by $\langle T^{(P)}_{ij} \rangle$~\cite{Lenny_99}. Apparently,
this stress tensor satisfies the trace Ward identity associated
with Lifshitz scaling symmetry~\cite{Marika_08},
  \be \label{ward_iden} z \,  \langle T^{0}_{0}\rangle + \langle
T^{i}_{i}\rangle =0 \, ,
 \ee
where at this order $\langle T^{0 (P)}_{0}\rangle =0$ by
construction and $\langle T^{i (P)}_{j}\rangle $ is traceless due
to the traceless polarization tensor $\xi_{ij}$. The conservation
of the energy-momentum is also trivially satisfied.

\subsection{Backreacted Geometry: Lifshitz Black Brane}
The gravitational waves will backreact on the metric, which in
turn gives corrections to the stress tensor. The second order
correction in terms of small boundary values of gravitational
waves $\phi$ is found to be positive, and is called the {\it
interest} in~\cite{Ford_99,Lenny_99}, with which to repay the
energy loan obtained in the first order perturbations. For a
specific choice of $\phi$ and the smearing function to be
introduced later, we find that the {\it interest} may possibly be
greater than the {\it principal} so that the sum of two pieces
becomes positive in accordance with the quantum interest
conjecture. We will argue later that the higher order terms,
{after being smeared over the parameter specifying
the prescribed path, are subleading to both the {\it principal}
and {\it interest}. So, one can obtain sensible results by keeping the
terms up to  second order in gravitational perturbations.}

The second order perturbations in principle can be calculated by
solving nonlinear field equations~(\ref{EoM}) order by order with
an input of the linear order results. In~\cite{Lenny_99},
Polchinski {\it et.al.} consider the spherically symmetric
gravitational wave as the first order perturbation, which is
generated at small $r$ in the bulk and then propagates toward the
boundary at larger $r_b$. According to the Birkhoff theorem for
cosmology constant~\cite{Kovalchuk_80}, any spherically symmetric
solution of source-less Einstein equations but with a cosmological
constant, is locally isometric to a region in Schwarzschid-de
Sitter(anti-de Sitter) spacetime characterized by a mass parameter
and cosmological constant. Thus, by means of the above-stated
theorem, the backreacted geometry near the boundary can be
parameterized by the metric of a neutral non-rotating AdS black
hole with the same total energy as that of gravitational waves due
to the spherically symmetric distribution of energy density.  As
for Lifshitz spacetime under consideration, the spherically
symmetric static black hole solutions for the theory with
action~(\ref{action}) are reviewed in~\cite{Bryn_09}. It is
possible to extend the Birkhoff theorem to include massive vector
fields as in the action (\ref{action}) and is discussed
in~\cite{Kovalchuk_80}. Since the gravitational wave perturbation
in~(\ref{graviton}) is found to propagate in Lifshitz spacetime
with spherical symmetry, along the lines of above arguments on AdS
space, we can then parameterize the backreacted metric by that of
a Lifshitz black hole with energy density contributed from
gravitational waves. As mentioned previously, the obtained second
order perturbations can be checked from straightforward
perturbation calculations and this deserves further study. Later
we mainly consider the $z=1$ case, which is pure AdS space,  to
explore the null energy conditions and their associated quantum
inequality.

The geometry near the boundary, which is relevant for the
calculations of stress tensor, is then asymptotic to the black
brane metric in Poincare-like coordinate as given
in~\cite{Peet_09}. Following~\cite{Peet_09}, we assume the asymptotic
form of the metric to be,
  \be
  \label{lifBH}
-\frac{r^{2z}}{L^{2z}}(1+I(r))dt^2+\frac{L^2}{r^2} (1+
B(r))dr^2+\frac{r^2}{L^2}(1+K(r))dx_idx_i \, ,
  \ee
and  perturbed vector fields are parameterized as
  \be
  \label{gauge pert}
  A_{\mu}=A\frac{r^z}{L^z}(1+J(r)+\frac12I(r))\delta^{0}_{\mu} \, .
  \ee
Their equations of motion can be found from
linearizing~(\ref{EoM}), and for $z\neq2$ they are
  \be \label{maxwell}
  -4zB(r)+3rI'(r)-z rB'(r)+2(3+z)rJ'(r)+2zrK'(r)+r^2I''(r)+2r^2J''(r)=0 \, ,
  \ee
  and
  \bea \label{einstein}
  &&-z (5+z) B(r)-2(z-1)(4+z)J(r)+(4+z)rI'(r)-zr B'(r)-2(z-1)rJ'(r) \nonumber \\
  && \quad\quad\quad\quad\quad\quad\quad\quad\quad\quad\quad\quad\quad\quad\quad\quad\quad\quad\quad\quad+2zrK'(r)+r^2I''(r)=0\, , \nonumber\\
  &&(4+z+z^2) B(r)+2(z-1)zJ(r)-(2+z)rI'(r)-(2+z) rB'(r)-2(z-1)rJ'(r) \,\nonumber \\
  &&\quad\quad\quad\quad\quad\quad\quad\quad\quad\quad\quad\quad\quad\quad\quad\quad\quad\quad\quad +6rK'(r)+r^2I''(r)+2r^2K''(r)=0 \, ,\nonumber \\
  &&-(4+z+z^2) B(r)+2(z-1)zJ(r)+zrI'(r)-rB'(r)+2(z-1)rJ'(r)\nonumber\\
  &&\quad\quad\quad\quad\quad\quad\quad\quad\quad\quad\quad\quad\quad\quad\quad\quad\quad\quad\quad\quad+(5+z)rK'(r)+r^2K''(r)=0
  \, , \eea
where the prime means the derivative with respect to $r$. Eq.(\ref{maxwell}) is given by Maxwell equations while the others~(\ref{einstein}) come
from Einstein equations.  According to~\cite{Peet_09}, the gauge is fixed by choosing
$B(r)=0$. In this case, the relevant solutions in the limit of
large $r$ for constructing the boundary energy-momentum tensor
later are obtained as
 \bea
 \label{bbrane}
 &&J(r)=-C\frac{z+1}{z-1}\,r^{-z-2} \, ,\nonumber\\
 &&I(r)=C\frac4{z+2}\, r^{-z-2} \, , \nonumber\\
 &&K(r)=C\frac2{z+2}\, r^{-z-2} \, .
 \eea
For $z=1$, since the spacetime reduces to pure AdS space where
${\cal A}$ in (\ref{para}) vanishes, the perturbation of the
vector field (\ref{gauge pert}) also vanishes. Then $I(r)$ and
$K(r)$ in the metric perturbations are straightforwardly given
by~(\ref{einstein}) with $z=1$. Using the prescriptions
(\ref{stress}) and (\ref{cfactor}), they leads to the boundary
stress tensor to second order in gravitational wave perturbations,
which is given by
  \bea
  \label{bh energy}
 &&16\pi G_4 \langle
 T^{(2)}_{00}\rangle=\frac{4\, C}{z\, L^{3+z}}(z-2)\, ,\nonumber\\
&&16\pi G_4 \langle
 T^{(2)}_{ij}\rangle=\frac{2\, C}{L^{3+z}}(z-2) \delta_{ij} \, ,
  \eea
where $C$ can be determined by matching the black brane mass
density to that of gravitational waves. As long as $z \neq 2$, the
leading order results~(\ref{bbrane}) in a large $r$ expansion give
nonvanishing stress tensor. For $z=2$, the stress tensor will be
determined by  their next order terms  that are beyond our
consideration in this paper~\cite{Peet_09}. Notice that different choices of the
gauge fixing may give different expressions for the metric, but
they all lead to the same stress tensor~\cite{Ross_09, Peet_09}.
In \cite{Peet_09},  the metric described by the solution
(\ref{bbrane}) is found, and is actually the Lifshitz black brane metric
with the energy density $M=\langle
 T_{00}\rangle$.  As in~\cite{Lenny_99}, we can also assume that all this energy density comes
 from the energy density of gravitational waves given by
  \bea
 M &=& \frac{L^{-3-z}}{8\pi G_4}\int dr \,
  r^{3+z}\bigg(\frac{r^{-2z-2}}{L^{-2z-2}}(\partial_t\phi)^2-(\partial_r\phi)^2\bigg)
  \nonumber\\
  &=& \frac{L^{-1+z}}{8\pi G_4}\int dr
  \, r^{1-z} \bigg((\partial_t\phi)^2-\phi\partial^2_t\phi\bigg) \,
  ,
  \eea
where the linearized equation of motion for the  $\phi$ field
(\ref{eom}) is used. Substituting the solution  (\ref{graviton})
to the above expression and using the orthogonal properties of
Bessel functions, the total energy density can be simplified as
  \be
  M=\frac{zL^{-3-z}}{2\pi G_4}\int_{0}^{\infty} \, d\omega
\,  \omega|\varphi(\omega)|^2 \,.
  \ee
Then the sub-leading boundary stress tensor, the {\it{interest}},
in terms of $M$ can be expressed as
   \be
   \label{st2}
\langle T^{(I)}_{00}\rangle =M,~~~~\langle
T^{(I)}_{ij}\rangle=\frac{z}2 M \delta_{ij}\,,
   \ee
where the trace Ward identity~(\ref{ward_iden}) is satisfied. All
components of the stress tensor in this order are uniform with no
spacetime dependence and the conservation of the obtained
energy-momentum  is apparent.

Based upon the idea of the bottom-up approach, {we assume} that
there exists a well defined field theory dual to this holographic
model. In the following, we will study the squeezed vacuum states
in a free field theory and find the counterparts of the {\it
principal} and {\it interest} of stress tensor. This suggests that
the squeezed vacuum state can be a possible candidate of the field
theory model dual to the gravity theory in gravitational wave
deformed Lifshitz geometry.

\section{Squeezed vacuum states in free field theory}
{In this section, the holographic stress tensor is
to be compared with the expectation value of stress tensor in free
field theory. This allows us to establish the dictionary between
the squeezing parameters of squeezed vacuum states for free fields
and the boundary values of gravitational waves.}
 The dual description
for squeezed vacuum states in bulk theory is assumed to be the
perturbed geometry from gravitational
waves~\cite{Lenny_99,yeh_16}. The faithful justification of the
duality needs the order by order comparison for all correlations
obtained from bulk theory and boundary field theory. However, here
we will restrict ourselves to the comparison up to second order in
an expansion of small squeezing parameters.

We now consider the squeezed vacuum states in 2+1-dimensional free
field theory with $N^2$ real scalars, $\Psi_{m n}$ where $m, n=1,2
...N$. It is known that the definition of the stress tensor is not
unique, and the so-called improved form of conformal invariance
can be constructed to be~\cite{bi_dav}
\be \label{st_full}
 T_{a b}= Tr  \bigg[ \frac{2}{3} \partial_{a}\Psi \partial_{b}
 \Psi-\frac{1}{6} \eta_{a b} (\partial_c \Psi )^2
 -\frac{1}{3} \Psi \partial_{a}\partial_{b} \Psi +\frac{1}{9}
 \eta_{a b} \Psi \partial^2 \Psi -\frac{1}{24} (\eta_{a b} \partial^2-\partial_{a} \partial_{b} ) \Psi^2 \bigg] \,
 \ee
with the metric of Minkowski spacetime $\eta_{a b}$. Later we will
evaluate the expectation value of the stress tensor with the
on-shell condition imposed. To simplify the notation, we write
each of the scalar fields as $\psi$, which can be expanded in
terms of creation and annihilation operators to be
\be \psi({\bf x},t)=\int \frac{d^2{\bf k}}{(2\pi)^2} \, ( a_{\bf k}
f_{\bf k}({\bf x},t) \, + \, a^{\dagger}_{\bf k} f^*_{\bf k}({\bf
x},t) )\, . \ee
 The above mode functions are chosen as
 \be f_{\bf k}
({\bf x},t)= e^{-i \omega_{k} t+i {\bf k} \cdot {\bf x}} \,
\label{modefun} \ee
with the dispersion relation $\omega_k=\vert {\bf k} \vert$. The squeezed vacuum states
are constructed out of the pure vacuum state by,
  \be
  |F \rangle=\exp\bigg[\frac12 \int d^2{\bf k}  \, \bigg( F^* ({\bf k})\,  a^{\dagger}_{\bf k} a^{\dagger}_{-\bf k}- F ({\bf k})\,  a_{\bf k} a_{-\bf k} \bigg) \bigg]|0\rangle \, ,
  \ee
where $F({\bf k})$ is the squeezing parameter. Thus, in the case
of weak squeezing, the  nonzero components of the expectation
value of the renormalized stress tensor can be found in a small
$F({\bf k})$ expansion as~\cite{Lenny_99}
  \begin{eqnarray}
  \langle T_{ii}^{(1)} \rangle_R  &=& N^2\int \frac{d^2\bf k }{(2\pi)^2}
  \frac{1}{2
\omega_k}\bigg(k_i^{{2}}-\frac12 {\omega_k}^2\bigg)\, F({\bf k}) \,
 e^{- 2 i  \omega_k
  t}+ {\rm h.c.} \, , \nonumber \\
  \langle T_{ij}^{(1)}
   \rangle_{R \, i \ne j }&=& N^2\int \frac{d^2\bf k }{(2\pi)^2}
  \frac{1}{2
\omega_k}k_i k_j\, F({\bf k}) \,
 e^{- 2 i  \omega_k
  t}+ {\rm h.c.} \, , \nonumber \\
  \langle T_{00}^{(2)}\rangle_R    &=& N^2 \int
  \frac{d^2{\bf k}}{(2\pi)^2} \, \frac{\omega_k}{2} \, |F({\bf k})|^2 \, , \nonumber
  \\
  \langle T_{ii}^{(2)}\rangle_R &=& N^2\int \frac{d^2{\bf k}}{(2\pi)^2} \frac{k_i^{{2}}}{2 \omega_k} \, |F({\bf k})|^2
  \, , \label{T_field}
  \end{eqnarray}
where the contributions from the pure vacuum state are subtracted.
Here we assume $F({\bf k})=F(-{\bf k})$ so that by construction
all nonzero components of the stress tensor have their
counterparts in the dual holographic model~\cite{Lenny_99}.

Now it is straightforward to find that $ \langle T^{(1)}_{00}
\rangle_R $ and  $ \langle T^{(1)}_{0i} \rangle_R $ vanish. The
behavior of $ \langle T^{(1)}_{ij} \rangle_R $ reveals  spatially
homogenous, but oscillatory in time. Thus, as compared with the
holographic calculation~(\ref{st1}), we identify this piece with
the {\it principal}. The stress tensor in second order $ \langle
T^{(2)} \rangle_R $  is positive and time independent. This piece
can be considered as the {\it interest} as in~(\ref{st2}). Notice
that both of the stress tensors have the same scaling as $N^2$.
Thus, we can relate $\varphi (k)$ in (\ref{st1}) with an
angle-averaged squeezing parameter of $F$ in~(\ref{T_field}) (
${\bar F}(k)=\frac1{2\pi}\int d\theta F({\bf k})$), up to a
numerical factor, as
  \be
  \label{dual}
    \varphi_{z=1}(k)= \frac{L^{3}}{\sqrt{2}} \,
    \sqrt{k} \, {\bar F}(k) \, .
  \ee
Note that the AdS/CFT dictionary, \be G_4=\frac{L^2}{N^2}
\label{G4} \ee is used. For a general Lifshitz scaling $z$, this
identification can be generalized as \be
  \label{dual2}
    \varphi(L^{z-1}k^z)=
    \frac{z}{\sqrt{2}}L^3k^{1-\frac{z}{2}} {\bar F}(k) \, .
  \ee
Once the relation between the squeezing parameter of quantum field
theory and the boundary values of gravitational waves in dual
holographic theory is established, we can then compare the
properties of stress tensors in strongly coupled field and free
scalar field theories. Notice that the above identification is not
uniquely specified for a general $z$. Nevertheless, by summing
over all momentum modes, different ways of identification will
still render the same expression of stress tensor given
by~(\ref{st1}) and~(\ref{st2}) in terms of dimensional quantities,
but with different overall constants that also depend on the
choices of the momentum-dependent function $F$.

\section{averaged null energy condition and quantum inequality}
We now study the averaged null energy conditions and quantum
inequalities given by a Lorentz invariant expression. Thus, the
strongly coupled quantum critical fields with $z=1$ will be
considered. The corresponding behaviors for free relativistic
fields will also be obtained for a comparison.

\subsection{strongly coupled fields with holographic approach}
It is quite straightforward to find whether or not the averaged
null energy condition~(\ref{ANEC2}) is satisfied in a strongly
coupled field theory for $z=1$. The null tangent vector is chosen
to be $K^{a}= (1,0,-1) $ and the trajectory along the null
direction is parameterized by an affine parameter $\lambda$ as
  $ x^{a} =\lambda \, K^{a} $.
Here we first consider the stress tensor of quantum critical
theories in dimension $d=3$ and for a general $z$.
 The squeezing parameter $F$ is assumed to be~\cite{Lenny_99}
   \be
  \label{sqp}
  F({\bf k})=\alpha\frac{k_2^2}{k^2}e^{-\sigma k^2}
\,  \ee where $\alpha$ is a small negative value. The width of the
{momentum-dependent function in the squeezing
parameter $F$} is characterized by $1/\sqrt{\sigma}$. The
angle-averaged $ F({\bf k})$ is
\be {\bar F} (k) = \frac{\alpha}{2} \, e^{-\sigma k^2}\, .
\label{a_sqp} \ee Then with the above specification of $\bar F$
and the identification of $\varphi$ in~({\ref{dual2}), the {\it
principal} and {\it interest} parts  of the stress tensor
in~(\ref{st1}) and (\ref{st2}) in terms of an affine parameter  $\lambda$
can be obtained straightforwardly as
 \be
 \label{prin}
\langle T^{(P)}_{22}\rangle=\frac{- \vert \alpha \vert
N^2}{64\sqrt{\pi}z}\sigma^{-\frac12-\frac1z}L^{\frac2z-2}\left(\frac12-\frac{\lambda^2}{\sigma}\right)e^{-\frac{\lambda^2}{\sigma}}
\, ,
 \ee
  \be
  \label{int}
  \langle T^{(I)}_{00}\rangle=\frac2{z}\langle
  T^{(I)}_{22}\rangle=\frac{3 \, \vert \alpha \vert^2 N^2}{32\pi
  z}2^{\frac1z-\frac72}\, \Gamma(\frac52-\frac1z)\, \sigma^{-\frac12-\frac1z}L^{\frac2z-2}
  \,
  \ee
  with  the chosen $\xi_{22}=1$.
Now we introduce the smearing function over the affine parameter to
be
\be g_{\lambda_0}= \frac{2}{\lambda_0}
\left(\frac{\lambda_0-|\lambda|}{\lambda_0}\right)^3 \, \big[ \,
\theta (\lambda+\lambda_0) +\theta (\lambda_0-\lambda) \, \big] \,
, \label{smear_fun} \ee which has support only on
$-\lambda_0<\lambda<\lambda_0$. The averaged value of null energy
density, expressed in a Lorentz invariant form for the case of
$z=1$, is given by
\bea \label{a_null_energy} && \int \, g_{\lambda_0} (\lambda) \,
\langle T_{a b} (\lambda) \rangle \, K^{a} K^{b} \, d \lambda
\nonumber \\
&&\quad\quad =
\frac2{\lambda_0}\int_{-\lambda_0}^{\lambda_0}d\lambda
\left(\frac{\lambda_0-|\lambda|}{\lambda_0}\right)^3\left(\langle
T_{00} \rangle +\langle T_{22}
  \rangle\right) (\lambda) \nonumber\\
  &&\quad\quad =\frac{3 \, |\alpha|^2 N^2}{32\pi}2^{-\frac32}\, \Gamma(\frac32)\, \frac{1}{\sigma^{\frac32}} - \frac{3 \, \vert \alpha
  \vert
N^2}{64\sqrt{\pi}}\frac{1}{\sigma^{\frac32}} \frac{\sigma}{\lambda_0^2}
\, ,
  \eea}
where $\langle T_{0i} \rangle =0$ is recalled  by  construction
from dual gravity theory. The averaged null energy density can be
negative for small $\lambda_0$, but satisfies the averaged null
energy condition in the large $\lambda_0$ limit. The {\it
interest} in~(\ref{st2}), which gives the positive contribution to
the stress tensor, is the second order term in a small squeezing
parameter $F$.  When $\lambda_0$ becomes large, the term of the
{\it interest}  becomes dominant  over the {\it principal}, obtained from~(\ref{st1}) by setting $ t=\lambda_0$ along the null
trajectory, which settles to zero in an oscillatory way. In the
end, the sum of two pieces of the stress tensor becomes positive
as $\lambda_0 \rightarrow \infty$, which is consistent with the
quantum interest conjecture. One may worry that the
perturbation expansion breaks down when the second order effect
becomes larger than its first order one. However we now show that
this is not the case. In general, the higher order terms in small
$F$ may depend on an affine parameter $\lambda$ or not. For those
$\lambda$-independent terms, their contributions to null energy
density is higher order in $ \alpha $ as compared to the first
term in~(\ref{a_null_energy}). As for the $\lambda$-dependent
terms, after being smeared over the affine parameter with the
function $g_{\lambda_0}$ ~(\ref{smear_fun}), their magnitudes
apparently increase with $\sigma$. Thus, for $\lambda_0 \gg
\sqrt{\sigma}$, the contributions from them to the null energy
density should depend on $(\sigma/\lambda_0^2)^{n}$ ($n\ge 1$) as
long as the null stress tensor such as~(\ref{prin}) and
(\ref{int}) is a regular function in the limit of $\lambda
\rightarrow 0$. Thus, the smeared higher order terms with
$\lambda$-dependence also lead to smaller contributions as
compared to the second term in~(\ref{a_null_energy}). Accordingly,
the results we obtain up to the second order terms are sensible.

To find a quantum inequality to constrain the magnitude and
duration of negative energy density based upon this holographic
setup, we will merely consider quantum critical fields in the
$z=1$ and $d=3$ case by following~\cite{kelly_14}. In the $z=1$
case, the background is just the 3+1-dimensional AdS space and the
metric can be written from (\ref{bmetric}) as
  \be
  ds^2=\frac{L^2}{\rho^2}(d\rho^2- du dv-dv du+dy^2) \, ,
  \ee
where we have set $\rho=\frac{L}{r}$,
$u=\frac1{\sqrt{2}L}(t+x_2)$, $v=\frac1{\sqrt{2}L}(t-x_2)$ and
$x_1=\frac{y}{L}$.
Then, the perturbed metric can be parameterized as
  \be
 \frac{L^2}{\rho^2}
 \big(d\rho^2-du dv-dv du+dy^2+\rho^3(t_{uu}du^2+t_{vv}dv^2+t_{uv}dudv+t_{yy}dy^2)\big)
 \, .
  \ee
In the limit of $r\rightarrow\infty$ ($\rho\rightarrow0$) and
using the identification (\ref{cfactor}), we can write
  \be \label{tuu}
  t_{uu}=\frac{8 \pi G_4 L}{3}  \, \langle T_{a b}\rangle K^{a} K^{b} +\mathcal{O}(\rho^2)\,
  ,
  \ee
where $K^{\mu}=(1, 0, -1)$ is described above. Since the
subleading terms of order $\rho^2$ in the metric expansion are
found to give small corrections in deriving quantum inequalities
below (see appendix for detailed discussions), we ignore them
henceforth. We now construct a causal (time-like) trajectory in
the bulk shown in Fig.1. The path is chosen to be close to the
boundary as $\varepsilon \rightarrow 0$, and  become a null
trajectory in the boundary for a large value of $u_0$.

\begin{figure}
\centering
\includegraphics[width=0.5\columnwidth,height=0.4\columnwidth]{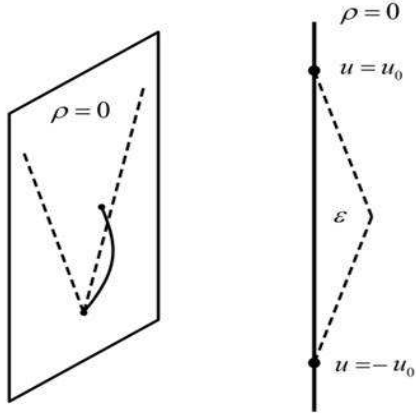}
\caption{{The time-like trajectory in the bulk by
following~\cite{kelly_14}.  }}
\end{figure}
The bulk causality gives an inequality with a bound,
  \be
  \label{bound5}
  \frac2{u_0}\int_{-u_0}^{u_0}d  u \left(\frac{u_0-|u|}{u_0}\right)^3\left(\langle T_{a b}(u,0) \rangle K^{a} K^{b}\right)\geq -\frac{3 \, N^2}{2\pi L^3}\frac1{u_0^{5/4}}
  \,
  \ee
that is consistent with~\cite{kelly_14}.  All detailed proof will be described in Appendix.
In the $u_0\rightarrow\infty$ ($\varepsilon\rightarrow0$) limit, since
the path in the boundary becomes null-like,  we can identify
$u=\frac{\sqrt{2}}{L}\lambda$, where $\lambda$ is an affine
parameter along the null geodesic.
The stress tensor is found to  satisfy
the average null energy condition as:
  \be
  \label{bound}
\lim_{\lambda_0 \rightarrow \infty}\,  \int  d \lambda \,
g_{\lambda_0} (\lambda) \, \left( \langle T_{a b} (\lambda)
\rangle K^{a} K^{b} \right)\ge 0 \,
  \ee
with the sampling function $g_{\lambda_0}
(\lambda)$~(\ref{smear_fun}). Thus, we have a general proof of the
null energy condition for the holographic stress tensor in $z=1$
case as the sampling function is chosen in~(\ref{smear_fun}).

Nevertheless  the above lower bound for a finite $\lambda_0$ is
not invariant under rescaling of an affine parameter that may not
be meaningful. In free relativistic field theory, such lower bound for
null-like geodesics does not exist either, except in $1+1$
dimensions as will be seen later.

Therefore we come to find a possible quantum inequality by
evaluating the null-contracted energy density along some other
path rather than a null geodesic. In~\cite{FE_03}, the
null-contracted energy density along the time-like path is
considered. Such a transverse smearing~\cite{wald}  gives a lower bound on the null energy density. To do
so, in the holographic framework, we evaluate $t_{uu}$ along a
time-like path in the boundary by setting $v=-\overline{v}(u)$ in
(\ref{causal}) (also see Appendix for details).
Thus, the quantum inequality  is achieved in terms of
$v=-\overline{v}(u)$ with a
 bound,
  \be
  \label{bound5_1}
  \frac2{u_0}\int_{-u_0}^{u_0}d  u \left(\frac{u_0-|u|}{u_0}\right)^3\left(\langle T_{a b}(u,-\overline{v}(u)) \rangle K^{a} K^{b}\right)\geq -\frac{3 \, N^2}{2\pi L^3}\frac1{u_0^{3/2}}
  \, .
  \ee
The prescribed time-like path has velocity $V \simeq 1+\frac{d
\overline{v}(u)}{du}\simeq 1-u_0^{-3/2}$}. In particular, for a
large value of $u_0$, the path is close to the null trajectory
with Lorentz factor $\gamma \simeq (\tau_0/L)^3 \gg1$ where  a proper time
$\tau_0 \simeq \gamma^{-1} u_0 \simeq L u_0^{1/4}$.
The bound can be expressed in a Lorentz invariant
form:
  \be
  \label{bound3}
  \int  d  \tau \, g_{\tau_0} (\tau)\, \left(\langle T_{a b}(\tau)
  \rangle K^{a} K^{b} \right)\gtrsim -\frac{3 \, N^2}{2\pi L^3}\frac{L^6}{\tau_0^6}
  \, ,
  \ee
that is now a function of the proper time with the smearing
function $g_{(\tau_0)} (\tau) $ given in~(\ref{smear_fun}). Again,
as $\tau_0 \rightarrow \infty$, the above inequality leads to the
averaged energy condition.
 Notice that since $V^a=(\gamma,0,-\gamma V)$ and
$K^a=(1,0,-1)$,  $(V_a K^a)^2\simeq \gamma^2 \simeq L^6/\tau_0^6 $
for a large $\gamma$ factor. Thus the right hand side of the bound
can be written in the form $(V_a K^a)^2$ of manifest invariance
under the rescaling of an affine parameter. This
inequality~(\ref{bound3}) is one of the main results in this work.
Notice that the above inequality is achieved by explicitly
constructing a particular path. From the boundary point of
view~\cite{Visser_09}, we may regard $L$ as a characteristic
length scaling of quantum critical theories. The  $1/L^3$
dependence in~(\ref{bound3}) gives correct units for energy
density in 2+1-dimensional spacetime. Later we will compare with
similar inequality obtained from free relativistic field theory in
that the expected dependence with units of energy density is
characterized by $1/\tau_0^3$. For $\tau_0 > L$, the inequality
from strongly coupled field will give a weaker constraint on
negative null energy density as compared with the constraint from
the free relativistic field when they both have same value of
$(K^a V_a)^2$.

To see the consequence of the negative lower bound, we will derive
the constraint on the duration of null negative energy. The averaged stress tensor over a time-like path can be
found from~(\ref{prin}) and~(\ref{int}) in that   the affine
parameter is replaced by a proper time, $\lambda \rightarrow \gamma \tau$ in the case of $z=1$.
We can see that,  within the initial time $\Delta \tau =\sqrt{ 2
\sigma}/\gamma$, the null energy density is dominated by the piece of
negative energy density with its magnitude in a spatial region
$\Omega_2$ to be
  \be \label{negE}
\frac{|\Delta E|}{\Omega_2}=\frac{|\alpha|N^2}{128 \sqrt{\pi} \,\,
\sigma^{3/2}}
  \, .
  \ee
With the result~(\ref{a_null_energy}) by
letting $\lambda =\gamma \tau_0$ and $z=1$,  the bound
(\ref{bound3}) gives an inequality  as
 \be
   \label{bound4}
\frac{N^2}{\sigma^{3/2}}\left(\frac{3 |\alpha|^2}{256 \sqrt{2}
\sqrt{\pi}}-\frac{3 | \alpha|}{64 \sqrt{\pi}
}\frac{\sigma}{\gamma^2 \, \tau_0^2}\right)\gtrsim -\frac{3 \,
N^2}{2\pi L^3}\frac{L^6}{\tau_0^6} \, .
   \ee
Within the above $\Delta \tau$, the most stringent bound imposed
on negative energy density can be obtained by ignoring its
positive contribution and letting $\sigma=\tau_0^2 \gamma^2
|\alpha|$, giving
 \be
 \label{bound6}
 \frac{|\Delta E|}{\Omega_2}(\Delta \tau)^{3}\le
  \frac{N^2}{\sqrt{2}\pi} \sqrt{|\alpha|} \,
  \frac{L^3}{\tau_0^3} \, .
  \ee
Again, the above bound is valid for $\tau_0 \gg L$ from the holographic
construction.
Recall that the above quantum
inequality is for the time-like trajectory with velocity
 \be \label{V}  V=1-\frac1{(\tau_0/L)^6} \, ,
 \ee
which means that the bound is valid only for the geodesic path very
close to be null. Later we will make a comparison with the quantum inequality
from free relativistic  field theory with the same $V$, where a more
restricted condition on negative energy is expected.

\subsection{free  field theory}
We now turn to considering the average null energy condition for conformally
coupled free massless scalar fields in $2+1$ dimensional Minkowski
spacetime for a comparison. We choose the small squeezing
parameter $F$ in (\ref{sqp}). With the same null tangent vector
$K_{a}=(1,0,-1)$ as above and the sampling function~(\ref{smear_fun}), the averaged
null energy for the stress tensor  over an affine parameter, which can be obtained from~(\ref{T_field}) by letting $\lambda=t$, up to the
second order in small $F$, becomes
\bea \label{a_null_energy_freefield} && \int \, g_{\lambda_0} (\lambda) \,
\langle T_{a b} (\lambda) \rangle_R K^{a} K^{b} \, d \lambda =
\frac2{\lambda_0}\int_{-\lambda_0}^{\lambda_0}d\lambda
\left(\frac{\lambda_0-|\lambda|}{\lambda_0}\right)^3\left(\langle
T_{00} \rangle_R +\langle T_{22}
  \rangle_R\right) (\lambda) \nonumber\\
  &&\quad\quad \quad\quad \quad\quad \quad\quad\quad\quad \quad\quad \quad\quad \quad\quad=\frac{9 \, \vert\alpha\vert^2 N^2}{1016 \sqrt{\pi}
  } \sigma^{-\frac32}- \frac{ 3 \, \vert \alpha
  \vert
N^2}{64\sqrt{\pi}}\sigma^{-\frac32}\frac{\sigma}{\lambda_0^2} \,
.
  \eea
We then find the associated quantum inequality, if exists. The
explicit expressions for the relevant $T_{a b}$ are given
by~(\ref{st_full}),
   \bea
   \label{tdd}
   T_{00}&=& N^2 \bigg( \frac{3}{4} \dot{\psi}^2 (x) -\frac{1}{4}
   \big(\dot{\psi}^2 (x) -(\nabla \psi (x))^2 \big) -\frac{1}{4}
   \psi (x) \ddot{\psi}(x) -\frac{1}{36} \psi (x) \partial^2 \psi (x)
   \bigg) \, , \nonumber\\
   T_{ii} &=&  N^2 \bigg( \frac{3}{4} (\partial_i {\psi} (x))^2 +\frac{1}{4}
   \big(\dot{\psi}^2 (x) -(\nabla \psi (x))^2 \big) -\frac{1}{4} \psi (x) \partial_i^2 \psi
   (x)+\frac{1}{36}  \psi (x) \partial^2 \psi (x) \bigg) \, ,
   \nonumber\\
    T_{0 i} &=& N^2 \bigg( \frac{2}{3}  \dot{\psi} \partial_i
 \psi
 -\frac{1}{3} \psi \partial_{i} \dot{\psi}  \bigg) \,.
\eea For the purpose of showing potential divergence in deriving
 quantum inequalities, let us consider the trajectory
$x^a_{\epsilon}= \lambda (1,0,-{\mathrm { v}}) $, where ${\mathrm{
v}}=1-\epsilon$, and in the limit $\epsilon \rightarrow 0^+$ the
trajectory is along the null geodesic. The mode
functions~(\ref{modefun}) evaluated along this nearly null
geodesic are
 \be f_{{\bf k}, {\mathrm v}}
(x)= e^{-i (k+ {\mathrm{ v}}  k_2) \lambda } \,.  \ee Now the mode
functions are smeared by the sampling function $g_{\lambda_0}(\lambda)$
in~(\ref{smear_fun}), and become:
\bea \label{h_pm}  h^{\pm}_{{\bf k} {\mathrm v}} &\equiv & h_{{\bf
k} {\mathrm v}}= \frac{2}{\lambda_0} \int_{-\lambda_0}^{\lambda_0}
\, d \lambda \, \Bigg( \frac{\lambda_0-\vert \lambda
\vert}{\lambda_0}
\Bigg)^3 \frac{ e^{\pm i (k+ {\mathrm{ v}} k_2) \lambda}}{ \sqrt{2 k}} \nonumber\\
&=& -   \sqrt{ \frac{2}{k}} \frac{1}{(k+ {\mathrm{ v}} k_2)^4
\lambda_0^4} \, \Big[ 12- 12 \, \cos[(k+ {\mathrm v} k_2)
\lambda_0]
  -6 (k+ {\mathrm{ v}} k_2)^2 \lambda_0^2 \Big] \, . \eea
  The dot means the derivative with respect to time $t$.
After taking integration by parts and dropping out irrelevant
total derivative terms, the null contracted  stress tensor, $
T_{ab} K^a K^b= T_{00}-T_{02}-T_{20}+T_{22}$,  can be cast into the
form of the product of the operator and its hermitian conjugate.
Thus, its expectation value  is positive definite as below:
   \bea
&& \int d\lambda\, g_{\lambda_0} (\lambda)  \langle T_{00}
-T_{02}-T_{20}+ T_{22} \rangle (\lambda) = N^2
\int \frac{d^2{\bf k}}{( 2\pi)^2} \, \frac{d^2{\bf k'}}{((2 \pi)^2} (k + k_2)  ( k' +k'_2 ) \, h_{{\bf k}
{\mathrm v}} h_{{\bf k'} {\mathrm v}} \nonumber\\
&& \quad\quad\quad\quad \quad\quad\quad \times \bigg[ -\langle a_{\bf k} \,
a _{\bf k'} + \langle a_{\bf k} \,
a^{\dagger}_{\bf k'} \rangle \, +\langle a^{\dagger}_{\bf k} \, a
_{\bf k'} \rangle \,  - \langle a^{\dagger}_{\bf k} \,
a^{\dagger}_{\bf k'} \rangle \, \bigg] \ge 0 \, .
   \eea
As such, the renormalized $\langle T_{ab} \rangle K^a K^b$ by
normal-ordering is found to have a lower bound,
  \bea \label{inq}  \int d\lambda \, g_{\lambda_0} (\lambda) \langle
T_{a b}(\lambda)\rangle_R  K^a K^b && \ge - N^2\int  \frac{d^2{\bf
k}}{( 2\pi)^2} \, \frac{d^2{\bf k'}}{((2 \pi)^2} (k + k_2)  ( k'
+k'_2 ) \, h_{{\bf k}
{\mathrm v}} h_{{\bf k'} {\mathrm v}} \nonumber\\
&& \quad \quad\quad\quad\quad\quad\quad \times  \bigg[
\left(\langle a_{\bf k'} \, a^{\dagger}_{\bf k} \rangle - \langle
a^{\dagger}_{\bf k} \, a_{\bf k'} \rangle \right) \bigg]   \nonumber\\
&&
 =  - N^2 \int \frac{d^2 {\bf k}}{(2
\pi)^2} (k+k_2)^2 \, h_{{\bf k}{\mathrm v}}^2 \, = -N^2 \,
\frac{27}{5} \, \frac{1 }{\lambda_0^3 \, (1+{\mathrm v}^2)
\,\sqrt{1-{\mathrm v}^2}} \,\nonumber\\ && \simeq -N^2 \,
\frac{27}{10} \, \frac{1}{\lambda_0^3 \epsilon}
 \eea
The integral diverges in the light-cone limit $\epsilon
\rightarrow 0^+$. Thus, for a finite $\lambda_0$, there is no
lower bound for free relativistic field theory. This is consistent with the findings
in~\cite{FE_03}.

According to~\cite{FE_03}, an alternative way to obtain  quantum
inequalities is to consider the null-contracted stress tensor in a
rest frame of the time-like trajectory described by a proper time
$\tau$. In this frame, the negative lower bound for the average
null energy is obtained from~(\ref{inq}) by setting ${\mathrm v}=0$ and
replacing $\lambda_0 \rightarrow \tau_0$. Then it can be written in
the form that is invariant under the rescaling of an affine
parameter,
\be  \int d\lambda \, g_{\tau_0} (\tau) \langle T_{a
b}(\tau)\rangle_R  K^a K^b   \ge  -N^2 \, \frac{27}{5} \,
\frac{(K^a V_a)^2}{\tau_0^3} \, , \ee where $V^a$ is a velocity
vector of the time-like trajectory. To compare with the quantum
inequality for strongly coupled fields in~(\ref{bound3}), we
substitute $V^a$ with velocity~(\ref{V}), namely $ (K^a V_a)^2
\approx (\tau_0/L)^6$, into above result. For $\tau_0 \gg L$, it
seems to indicate that the quantum inequality for strongly coupled
fields gives weaker constraint than the one for free relativistic fields.

To find the constraint on the negative null energy, we choose the
small squeezing parameter $F$ as in (\ref{sqp}) and following the
procedure above, the magnitude of the negative  null energy  within the duration  time  $\Delta \tau$ obeys an
inequality,
  \be  \frac{\vert \Delta E \vert }{\Omega_2}  (\Delta \tau )^3 <
\frac{9 N^2}{ 5\sqrt{2}}  \sqrt{|\alpha|} \, (K^a V_a)^2 \, ,\ee
where $(K^a V_a)^2 \simeq (L/\tau_0)^6 \ll 1$. Thus, as compared
with (\ref{bound6}), we find that in strongly coupled field theory
larger  $\Delta \tau$ is allowed for a fixed $\vert \Delta E
\vert$.  Whether or not the above feature is also true for other
quantum states with negative energy density needs the further
study.

\section{Conclusion}
In this paper, we have used the AdS/CFT correspondence to study
the expectation value of stress tensor in $2+1$-dimensional
strongly coupled quantum critical theories with Lifshitz scaling
symmetry. The holographic dual theory is the gravity theory in
3+1-dimensional Lifshitz geometry. We adopt the consistent scheme
to construct the boundary stress tensor that satisfies the trace
Ward identity due to Lifshitz scaling symmetry. The boundary
stress tensor, obtained from the gravitational wave deformed
Lifshitz geometry, can be found up to second order in
gravitational wave perturbations, and is compared to its
counterpart for free scalar fields at same order in an expansion
of small squeezing parameters. We can then relate the boundary
values of gravitational waves to the squeezing parameters of
squeezed vacuum states. We find that, in both cases with $z=1$,
the leading order term of stress tensor for a small squeezing
parameter  reveals oscillation between negative and positive
values in time and settles to zero at late times. This piece can
be identified with the {\it principal} based upon the quantum
interest conjecture in \cite{Ford_99}.  The subleading term is
spacetime uniform with a positive constant value, and can be
identified with the {\it interest}. The averaged null energy
condition along the null trajectory with an affine parameter
$\lambda_0$ is studied in the $z=1$ case. We find that, although
the sum of the leading and subleading terms in a small squeezing
parameter expansion may be negative for small $\lambda_0$, the
averaged null energy density then becomes positive in the limit of
$\lambda_0 \rightarrow \infty$, which satisfies the averaged null
energy condition, and is thus consistent with the quantum interest
conjecture~\cite{Ford_99}.

We then try to find the associated negative lower bound to the
averaged null-contracted stress tensor, also  for the case of
$z=1$. For the average over the complete null geodesic, the bound
exists for quantum critical fields while there is no such bound  for free relativistic
fields. An alternative smearing is
introduced, in that the null-contracted stress tensor is averaged
over time-like trajectories. For the trajectories along nearly
null directions, the bounds are found. We find the weaker
constraint on the magnitude and duration of negative null energy
density in strongly coupled field theories as compared with the
constraint in free relativistic field theories. Thus, it implies
that the negative energy density for strongly coupled fields
 might be allowed to last longer, before the
overcompensation from the piece of positive energy density, than
for weakly coupled fields. This might give some
implications to the existence of exotic spacetimes sourced by
quantum fields with negative energy density. Whether or not this
feature  remains true in more general time-like trajectories,
spacetime dimensions and other quantum states deserves future
study.
\section{Appendix}
In this Appendix, we will present the detailed description on how
the inequalities~(\ref{bound5}) and~(\ref{bound5_1}) are obtained.
In the $z=1$ case, the background is just the 3+1-dimensional AdS
space and the metric can be rewritten from (\ref{bmetric}) as
  \be
  ds^2=\frac{L^2}{\rho^2}(d\rho^2- du dv-dv du+dy^2) \,
  \ee
 with the chosen coordinates $\rho=\frac{L}{r}$,
$u=\frac1{\sqrt{2}L}(t+x_2)$, $v=\frac1{\sqrt{2}L}(t-x_2)$ and
$x_1=\frac{y}{L}$.
Then, the perturbed metric can be parameterized as
  \be
 \frac{L^2}{\rho^2}
 \big(d\rho^2-du dv-dv du+dy^2+\rho^3(t_{uu}du^2+t_{vv}dv^2+t_{uv}dudv+t_{yy}dy^2)\big)
 \, .
  \ee
In the limit of $r\rightarrow\infty$ ($\rho\rightarrow0$) and
using the identification (\ref{cfactor}), we can write
  \be
  \label{tuu2}
  t_{uu}=\frac{8 \pi G_4 L}{3}  \, \langle T_{a b}\rangle_R K^{a} K^{b} +\mathcal{O}(\rho^2) \,
  ,
  \ee
where $K^{a}=(1, 0, -1)$ is described above. Other components of
metric perturbations can have the similar expansion, but here we
will particularly focus on the $t_{uu}$ component. We now
construct a causal (time-like) trajectory in the bulk shown in
Fig.1 given by
   \bea
   \label{causal}
   &&\bar{\rho}(u)=\varepsilon\left(\frac{u_0-|u|}{u_0}\right)\, ,\nonumber\\
   &&\bar{v}(u)=\frac{\varepsilon^{2-\Delta}}{2u_0}\left(\frac{u_0+u}{u_0}\right)+\frac{\varepsilon^3}{2}\int_{-u_0}^{u}du'\left(\frac{u_0-|u'|}{u_0}\right)^3 t_{uu} (u',v(u'))
   \eea
for $-u_0\le u\le u_0$ and $\Delta>0$. In addition,  $u=\pm u_0$ are on the
boundary. We require $\varepsilon$ to be small so the path is
close to the boundary, and $u_0$ to be large for having an
interesting bound. The (bulk) time-like condition
$g_{\mu\nu}\frac{dx^{\mu}}{du}\frac{dx^{\nu}}{du}\le 0$ gives
   \be
   (\bar{\rho}')^2-2\bar{v}'+\bar{\rho}^3\left(t_{uu}+t_{uv}\bar{v}'+t_{vv}\bar{v}'^2\right)\le
   0 \, .
   \ee
The prime here denotes the derivative with respect to $u$. With
the prescribed path, the bulk causality above is satisfied if
  \be
  \label{bulkcau}
 \frac{\varepsilon^{2-\Delta}}{u_0^2}>\mathcal{O}(\varepsilon^6) \, .
  \ee
Having described the causal trajectory (\ref{causal}), the theorem
in \cite{Wald_00} shows that the end point located at the boundary
$\rho=0$ must be also causally separated,
  \be
  \label{bineq}
\bar{v}(u_0)-\bar{v}(-u_0)=\frac{\varepsilon^3}{2}\int_{-u_0}^{u_0}du'\left(\frac{u_0-|u'|}{u_0}\right)^3t_{uu}+\frac{\varepsilon^{2-\Delta}}{u_0}\geq
0 \, ,
  \ee
leading to the quantum inequality for the null energy
density. Along the trajectory (\ref{causal}), the value of the
coordinate $v$ is small, and $t_{uu}$ can be expanded in terms of
small $v$.
 Thus the inequality (\ref{bineq}) gives,
    \be
    \label{nulliniq1}
   \frac2{u_0}\int_{-u_0}^{u_0}du'\left(\frac{u_0-|u'|}{u_0}\right)^3\left(t_{uu}(u',0)+\partial_v
   t_{uu}(u',0)\bar{v}(u')+...\right)\geq
   -\frac{4\varepsilon^{-1-\Delta}}{u_0^2} \, .
   \ee
   Assuming that the integral over energy-momentum tensor and all its
derivatives are all finite~\cite{kelly_14}, we may ignore the sub-leading
terms if
  \be
  \label{bdycau}
  \frac{\varepsilon^{-1-\Delta}}{u_0^2}>\mathcal{O}(\bar{v}) \, .
  \ee
Then the inequality (\ref{nulliniq1}) becomes
 \be
 \label{qe}
   \frac2{u_0}\int_{-u_0}^{u_0}du'\left(\frac{u_0-|u'|}{u_0}\right)^3 t_{uu}(u',0)\geq
   -\frac{4\varepsilon^{-1-\Delta}}{u_0^2} \, .
  \ee
To satisfy (\ref{bulkcau}) and (\ref{bdycau}), we find that the
most stringent bound is obtained when considering
$\mathcal{O}(\bar{v}) \sim \varepsilon^3 u_0$, and  letting
$\Delta\rightarrow 0$ and $u_0\rightarrow \varepsilon^{-4/3}$. The
inequality~(\ref{bound5}) is obtained. Note that in finding
quantum inequalities, we ignore the order $\rho^2$ terms in
(\ref{tuu2}) in the form of $\rho^2 s_{uu}$ where $s_{uu}$ is the
algebraic function of  the renormalized expectation value of
stress tensor and its derivatives~\cite{kelly_14}. Nevertheless
for a bounded renormalized expectation value of stress tensor such
as the one in~(\ref{prin}), and (\ref{int}), the smeared $t_{ a b}
$ over $u$ is in general independent of $u_0$ or inversely
proportional to the powers of $u_0$. The corrections from the
ignored terms, when letting $u_0\rightarrow \varepsilon^{-4/3}$,
are at most of order $\varepsilon^2$, which give ignorable
contributions as compared with the lower bound in the right hand
side of (\ref{qe}) in the limit of $\varepsilon \rightarrow 0$.

As for the average over the time-like path,  we evaluate $t_{uu}$
along a path in the boundary by setting $v=-\overline{v}(u)$ in
(\ref{nulliniq1}). Then, in this case the most stringent bound
which is solely due to (\ref{bulkcau}), is by setting
$\Delta\rightarrow 0$ and $u_0\rightarrow \varepsilon^{-2}$ to
achieve~(\ref{bound5_1}).

\begin{acknowledgments}
We would like to thank the authors of~\cite{kelly_14} with their
permission to use the graph. This work was supported in part by
the Ministry of Science and Technology, Taiwan.
\end{acknowledgments}


\begin{thebibliography}{999}
\bibitem{PENROSE_65}
R. Penrose, "Gravitational collapse and space-time singularities",
Phys. Rev. Lett. {\bf 14}, (1965) 57 (1965); S. Hawking, "The
Occurrence of singularities in cosmology", Proc. Roy. Soc. Lond. A
{\bf 294}, 511 (1966);
 S. Hawking and R. Penrose, "The Singularities
of gravitational collapse and cosmology", Proc. Roy. Soc. Lond. A
{\bf 314}, 529  (1970).
\bibitem{EP}
H. Epstein, V. Glaser, and A. Jaffe, "Nonpositivity of energy
density in quantized field theories", Nuovo Cimento {\bf 36}, 1016
(1965).

\bibitem{FO}
L. H. Ford, "Quantum coherence effects and the second law of
thermodynamics", Proc. R. Soc. London A {\bf 346}, 227 (1978).

\bibitem{DA}
P. C. W. Davies, Phys. Lett. B {\bf 113}, 393 (1982).

\bibitem{FOR}
L. H. Ford and T. A. Roman, Phys. Rev. D {\bf 41}, 3662 (1990) ; L.
H. Ford and T. A. Roman, Phys. Rev. D {\bf 46}, 1328 (1992).


\bibitem{MO}
M. S. Morris and K. S. Thorne, Am. J. Phys. {\bf 56}, 395 (1988); M.
S. Morris, K. S. Thorne, and U. Yurtsever, Phys. Rev. Lett. {\bf
61}, 1446 (1988).

\bibitem{AL}
M. Alcubierre, "The Warp drive: hyperfast travel within general
relativity", Class. Quantum Grav. {\bf 11}, L-73 (1994).

\bibitem{FOR_96}
L. Ford and T. A. Roman, "Quantum field theory constraints
traversable wormhole geometries",  Phys. Rev. D {\bf 53},  5496 (1996),
arXiv:gr-qc/9510071.




\bibitem{VISSER_98}
D. Hochberg and M. Visser, "The null energy condition in dynamic
wormholes", Phys. Rev. Lett. {\bf 81},  746 (1998),
arXiv:gr-qc/9802048;  M. Visser, S. Kar, and N. Dadhich,
"Traversable wormholes with arbitrarily small energy condition
violations", Phys. Rev. Lett. {\bf 90}, 201102  (2003),
arXiv:gr-qc/0301003.

\bibitem{ORI}
A. Ori, "Must time machine construction violate the weak energy
condition?", Phys. Rev. Lett. {\bf 71},  2517 (1993).

\bibitem{Myrzakulov}
R. Myrzakulov, L. Sebastiani,
S. Vagnozzi, and S. Zerbini,  "Static spherically symmetric solutions in
mimetic gravity: rotation curves and
wormholes", Class. Quantum Grav. {\bf 33}, 125005 (2016), arXiv:gr-qc/1510.02284.

\bibitem{FO1}
L. H. Ford, "Constraints on negative-energy fluxes", Phys. Rev. D
{\bf 43}, 3972 (1991).

\bibitem{For_95}
L. H. Ford and Thomas A. Roman, " Averaged energy conditions and
quantum inequalities",  Phys. Rev. D {\bf 51}, 4277 (1995).

\bibitem{PF}
M. J. Pfenning and L. H. Ford, Phys. Rev. D {\bf 57}, 3489 (1998);
L. H. Ford, M. J. Pfenning, and T. A. Roman, Phys. Rev. D {\bf 57},
4839 (1998).

\bibitem{FE}
C. Fewster and S. Eveson, Phys. Rev. D {\bf 58}, 084010 (1998).

\bibitem{FE_03}
 C. J. Fewster and T. A. Roman, "Null energy conditions in quantum field
 theory", Phys.  Rev.  D {\bf 67}, 044003 (2003).


\bibitem{Ford_99}
 L. Ford, T. Roman,
 ``The quantum interest conjecture'', Phys. Rev.  D {\bf 60},  104018 (1999), arXiv:gr-qc/9901074.


\bibitem{HWL}
J.-T. Hsiang, T.-H. Wu, and D.-S, Lee, Phys. Rev. D {\bf 77}, 105021
(2008).

\bibitem{Lee_12}
J.-T. Hsiang, T.-H. Wu, and D.-S. Lee, Ann. Phys. {\bf 327}, 522
(2012).

\bibitem{HS}
J.-T. Hsiang and L. H. Ford,  Phys. Rev. Lett. {\bf 92}, 250402
(2004).

\bibitem{Lee_06}
 J.-T. Hsiang and D.-S. Lee, Phys. Rev. D {\bf 73}, 065022 (2006).


\bibitem{AdSCFT} J.~M.~Maldacena, Adv.\ Theor.\ Math.\ Phys.\ {\bf 2}, 231 (1998)
(Int.\ J.\ Theor.\ Phys.\ {\bf 38}, 1113 (1999)); S.~S.~Gubser,
I.~R.~Klebanov and A.~M.~Polyakov, Phys.\ Lett.\ B {\bf 428}, 105
(1998); E.~Witten, Adv.\ Theor.\ Math.\ Phys.\ {\bf 2}, 253 (1998).


\bibitem{Herzog:2006gh}
C.~P. Herzog, A.~Karch, P.~Kovtun, C.~Kozcaz, and L.~G. Yaffe,
``Energy loss of a heavy quark moving through $\mathcal{N}=4$ supersymmetric Yang-Mills plasma'',
 JHEP {\bf 07}, 013 (2006).


\bibitem{Gubser_06}
S. S. Gubser, ``Drag force in AdS/CFT'', Phys. Rev. D \textbf{74}, 126005 (2006).

\bibitem{Teaney_06}
J. Casalderrey-Solana and D.~Teaney, ``Heavy quark diffusion in strongly
coupled $\mathcal{N}=4$ Yang-Mills theory'', Phys. Rev. D \textbf{74},
085012 (2006).

\bibitem{Son:2009vu}
D.~T. Son and D.~Teaney,``Thermal noise and stochastic strings in AdS/CFT'', JHEP {\bf 07}, 021 (2009).

\bibitem{Giecold:2009cg}
G.~C. Giecold, E.~Iancu, and A.~H. Mueller,``Stochastic trailing string and Langevin dynamics from AdS/CFT'', JHEP {\bf 07}, 033 (2009).



\bibitem{CasalderreySolana:2009rm}
J.~Casalderrey-Solana, K.-Y. Kim, and D.~Teaney,``Stochastic string motion above and below the world sheet horizon'', JHEP {\bf 12}, 066 (2009).

\bibitem{Huot_2011}
 S. Caron-Huot, P. Chesler and D. Teaney,
``Fluctuation, dissipation and thermalization in non-equilibrium $AdS_5$ black hole geometries'', Phys. Rev. D \textbf{84}, 026012 (2012).

\bibitem{Holographic QBM}
 J. Boer, V. Hubeny, M. Rangamani and M. Shigemori, ``Brownian motion in AdS/CFT'', JHEP \textbf{07}, 094 (2009); V. Hubeny and M. Rangamani,``A holographic view on physics out of equilibrium'', Adv. High Energy Phys. {\bf 2010}, 297916 (2010).

\bibitem{Tong_12}
 D. Tong and K. Wong,
``Fluctuation and dissipation at a quantum critical point'', Phys. Rev. Lett. {\bf 110}, 061602 (2013).

\bibitem{yeh_14}
C.-P. Yeh, J.-T. Hsiang, D.-S. Lee, "Holographic approach to nonequilibrium dynamics of moving mirrors coupled to quantum critical theories",
Phys. Rev. D {\bf 89}, 066007 (2014).

\bibitem{yeh_15}
C.-P. Yeh, J.-T. Hsiang, D.-S. Lee,"
Holographic influence functional and its application to decoherence induced by quantum critical theories",
Phys. Rev. D {\bf 91}, 046009 (2015).

\bibitem{yeh_16}
C.-P. Yeh, D.-S. Lee, "Subvacuum effects in quantum critical theories from holographic approach", Phys. Rev. D {\bf 93}, 126006 (2016).
 arXiv:1410.7111.


\bibitem{bousso_16}
R. Bousso, Z. Fisher, J. Koeller, S. Leichenauer, and A. C. Wall,
``Proof of the quantum null energy condition``,
 Phys. Rev. D {\bf 93}, 024017 (2016),  arXiv:gr-qc/1509.02542.

\bibitem{koeller_15}
J. Koeller, S. Leichenauer, ``Holographic proof of the quantum
null energy condition``,  arXiv:gr-qc/1512.06109.


\bibitem{kelly_14}
 W. R. Kelly and A. C. Wall,
``A holographic proof of the averaged null energy condition'', Phys. Rev. D {\bf 91}, 069902  (2015), arXiv:gr-qc/1408.3566;
Erratum: Holographic proof of the averaged null energy condition
[Phys. Rev. D 90, 106003 (2014)], Phys. Rev. D {\bf 91}, 069902(E) (2015).


 \bibitem{Lenny_99}
 J. Polchinski, L. Susskind and N. Toumbas,
 ``Negative energy, superluminosity and holography'',  Phys. Rev. D {\bf 60}, 084006 (1999), arXiv:hep-th/9903228.






\bibitem{Kachru_08}
 S. Kachru, X. Liu and M. Mulligan,
``Gravity duals of Lifshitz-like fixed points'', Phys. Rev. D {\bf
78}, 106005 (2008).

\bibitem{Marika_08}
 M. Taylor,
 ``Non-relativistic holography'',
 arXiv:hep-th/0812.0530; "Lifshitz holography",  arXiv:hep-th/1512.03554.

\bibitem{Kraus_99}
 V. Balasubramanian and P. Kraus,
 ``A stress tensor for anti-de Sitter gravity'', Commun. Math. Phys. {\bf 208}, (1999) 413-428  (1999),  arXiv:hep-th/9902121.

\bibitem{Myers_99}
 R. Myers,
 ``Stress tensors and casimir energies in the AdS/CFT corresponsce'', Phys. Rev.  D {\bf 60}, 046002 (1999),  arXiv:hep-th/9903203.


\bibitem{Marolf_99}
 S. Hollands, A. Ishibashi and D. Marolf,
 ``Counter-term charges generate bulk symmetries'', Phys. Rev. D {\bf 72}, 104025 (2005), arXiv:hep-th/0503105.

\bibitem{Ross_09}
S. Ross and O. Saremi,
 ``Holographic stress tensor for non-relativistic theories'',  JHEP {\bf 09},  009 (2009),   arXiv:hep-th/0907.1846.

\bibitem{Witt_09}
K. Schleich and  D. Witt,
 ``A simple proof of Birkhoff's theorem for cosmological constant'', J. Math. Phys.~{\bf 51}, 112502 (2010),
  arXiv:gr-qc/0908.4110.

\bibitem{Bryn_09}
E. Brynjolfsson, U. Danielsson, L. Thorlacius and T. Zingg,
 ``Holographic superconductors with Lifshitz scaling'', J. Phys. A: Math. Theor. {\bf 43}, 065401 (2010),
  arXiv:hep-th/0908.2611.


\bibitem{Kovalchuk_80}
K. Bronnikov and M. Kovalchuk,
 ``On a generalisation of Birkhoff¡¦s theorem '',  J. Phys.
A: Math. Gen. {\bf 13}, 187 (1980).




\bibitem{Peet_09}
 G. Bertoldi, B. Burrington and A. Peet,
 ``Thermodynamics of black branes in asymptotically Lifshitz spacetimes '', Phys. Rev.  D {\bf 80},  126004 (2009), arXiv:hep-th/0907.4755.

\bibitem{bi_dav}
N. D. Birrell, P. C. W. Davies, {\it Quantum Fields in Curved Space}
(Cambridge University Press).




\bibitem{wald}
E. Flanagan and R. M. Wald, Phys. Rev. D {\bf 54}, 6233 (1996).


\bibitem{Visser_09}
M. Visser, Phys. Rev. D {\bf 80}, 025011 (2009),  arXiv:hep-th/0902.0590.


\bibitem{Wald_00}
S. Gao and R. M. Wald, ``Theorems on gravitational time delay and
related issues'',  Class. Quant. Grav. {\bf 17},  4999 (2000),
arXiv:gr-qc/0007021.

\end{thebibliography}
\end{document}